# Enhanced Probing of Fermion Interaction Using Weak Value Amplification

Alex Hayat[†,1,2], Amir Feizpour[2], and Aephraim M. Steinberg[2]

[1]*Department of Electrical Engineering, Technion, Haifa 32000, Israel*

[2]*Centre for Quantum Information and Quantum Control, Department of Physics, University of Toronto, 60 St. George Street, Toronto ON, M5S 1A7, Canada*

We propose a scheme for enhanced probing of an interaction between two single fermions based on weak-value amplification. The scheme is applied to measuring the anisotropic electron-hole exchange interaction strength in semiconductor quantum dots, where both spin and energy are mapped onto emitted photons. We study the effect of dephasing of the probe on the weak-value-enhanced measurement. We find that in the limit of slow noise, weak-value amplification provides a unique tool for enhanced-precision measurement of few-fermion systems.

---

[†] alex.hayat@ee.technion.ac.il

Weak measurement is a concept in the study of quantum systems, based on obtaining information both from the initial state of the system and from the subsequent post-selection. Lately, this approach has been employed to gain new insights into the fundamental principles of quantum mechanics [1,2,3,4,5], as well as to enhance quantum metrology [6,7]. A measurement probe coupled weakly to a quantum system can obtain very little information from it, while the disturbance of the measurement onto the system remains negligible. If the system is prepared in an initial state $|i\rangle$ and post-selected in a final state $|f\rangle$, the mean size of the effect an ensemble of such systems would have on a measurement of the observable $A$ is given by $\langle A \rangle_w = \langle f|A|i\rangle / \langle f|i\rangle$, which is called the "weak value". An important property of these weak values is the fact that they do not necessarily lie within the eigenvalue spectrum of the observable $A$. In particular, for small overlap $\langle f|i\rangle$, the weak value can be significantly larger than the typical eigenvalues of $A$, resulting in "weak-value amplification" (WVA) [8].

Any practical measurement is subject to noise and the signal-to-noise ratio (SNR) is of paramount concern in any amplification scheme. Recently, WVA was shown to enhance the SNR in the presence of low-frequency noise [9]. All previous realizations of WVA were done with bosons (specifically photons). However, since many bosons can be prepared in the same state, it is always possible, in principle, to perform the measurement with a large number of bosons to obtain a high SNR. On the other hand, for fermionic systems the Pauli Exclusion Principle (PEP) makes it impossible to prepare more than one particle per mode. Therefore measurements on these systems have to be performed one fermion at a time, making them prone to slow noise. The advantages of WVA, therefore, may prove uniquely powerful for measurements on fermions.

Spin is an internal quantum property of particles with relatively weak coupling to other degrees of freedom, making it a good candidate for solid-state quantum information processing [10,11,12] and other applications such as spintronics [13]. In order to measure the weak coupling of a spin to other physical quantities, the ideas of WVA can be used to provide a unique method of obtaining a sensitive spectroscopic signature. The WVA approach can yield especially significant enhancement for systems that have to be studied on an individual basis such as semiconductor quantum dots (QDs).

Here we propose a WVA method for probing the spin-dependent energy splitting of a fermionic system. As a concrete practical example of enhanced probing of a fermion spin interaction, we consider the electron spin in a QD with the energy levels split by the anisotropic electron-hole exchange interaction energy, $\Delta E$ [14], and an energy level broadening (FWHM) of $\Gamma = \hbar/T_1$ where $T_1$ is the radiative decay time (Fig. 1). The WVA is used to deduce the anisotropic electron-hole exchange interaction strength, where the "system" and the "probe" of the measurement, the spin of the electron and its energy spectrum respectively, are prepared in a coupled state when the "system" is initialized. This preparation is different from the conventional description of WVA [8,9]; nevertheless we show that in this case the probe behavior is exactly equivalent to the more familiar case in certain regimes. We are interested in the case where $\Delta E < \Gamma$, and hence is unresolved. This makes direct probing difficult and is precisely the regime in which the interaction can be considered "weak" and the advantages of WVA can be brought to bear on the problem. This regime is especially interesting because it is used for QD-based entangled photon sources in order to erase the which-path information [12].

Usually the splitting is probed by pumping the quantum dot to the biexciton level and detecting the spectrum of the emitted photons, which includes the spectrum of the biexciton and the exciton decays. We propose to use the polarization of the emitted photons to initialize and post-select the system (i.e. the spin), and then to use the modified spectrum of the emitted photon to read out an enhanced energy shift proportional to the splitting. Therefore, the electron spin is preselected in a given initial state by projecting the first photon emitted in the biexciton cascade in the $XX^0$ line onto a given polarization, so that the QD in the $X^0$ state is prepared in a given spin superposition (Fig. 1 a). When $\Delta E < \Gamma$ the weakness criterion for weak measurement is met [8]. Once the excited state decays, the spin (energy) state of the QD is mapped onto the polarization (spectrum) of the emitted photon and the weak measurement of the spin "system" is then completed by detecting the spectrum of the emitted photon from the $X^0$ line – post-selected on a certain polarization (Fig. 1 b). The symmetric QD state $|S\rangle$ is mapped onto the horizontal photon polarization $|H\rangle$ (along the major axis of the QD) and the antisymmetric $|A\rangle$ state is mapped onto vertical polarization $|V\rangle$ (along the minor axis of the QD) [14] as shown in Table 1:

| QD angular momentum, energy | Photon polarization, energy |
|---|---|
| $|S\rangle = \frac{1}{\sqrt{2}}(|\uparrow\Downarrow\rangle + |\downarrow\Uparrow\rangle)$, $E_H = E_0 - \Delta E/2$ | $|H\rangle$, $E_H = E_0 - \Delta E/2$ |
| $|A\rangle = \frac{1}{\sqrt{2}}(|\uparrow\Downarrow\rangle - |\downarrow\Uparrow\rangle)$, $E_V = E_0 + \Delta E/2$ | $|V\rangle$, $E_V = E_0 + \Delta E/2$ |

Table 1. Mapping of the QD states to photon states. $|\uparrow\rangle$ and $|\downarrow\rangle$ denote electron states with angular momentum $J_z = +1/2$ and $J_z = -1/2$. $|\Uparrow\rangle$ and $|\Downarrow\rangle$ denote heavy hole states with angular momentum $J_z = +3/2$ and $J_z = -3/2$, where z is the growth direction

The energy spectrum of the photon is converted to position on a screen using a spectrometer and the measurement can be modeled by the effective Hamiltonian $H = (\eta \Delta E) S_z p_y$, where $S_z$ is the spin operator, $p_y$ is the transverse momentum on the spectrometer screen and $\eta$ is a constant given by the spectrometer geometry. By preparing and post-selecting the spin in appropriate states, one can more precisely determine the system-pointer coupling, $\Delta E$, by enhancing the SNR.

A radiative decay of a QD biexciton can occur in two paths: both the $X^0$ and $XX^0$ line photons emitted with H polarization, or both lines with V polarization [12]. The H-path and the V-path occur via different QD angular momentum states with different energies, split by the interaction under study here : $E_0 \pm \Delta E/2$. Therefore, if an H photon is emitted in the first step of the cascade (in the $XX^0$ line), the QD will be in the symmetric angular momentum state $|S\rangle$ with energy $E_H = E_0 - \Delta E/2$ (Table 1), whereas if a V photon is emitted, the QD is in the antisymmetric state $|A\rangle$ with energy $E_V = E_0 + \Delta E/2$ (Fig. 1). If the photon from the first step of the cascade is detected in a superposition polarization state $|i\rangle = \frac{1}{\sqrt{2}}\big[|H\rangle + |V\rangle\big]$ at t=0, the resulting QD state will be an energy-angular-momentum entangled state. In the following radiative decay of the exciton to the ground state on the $X^0$ spectral line, the emission from the $|S\rangle$ state results in an H photon, and central energy $E_0 - \Delta E/2$, while the $|A\rangle$ state results in a V photon with $E_0 + \Delta E/2$ central energy. In the decay of an entangled exciton state, therefore, the emitted photon will also be an entangled state:

$$|\Psi_0\rangle = \frac{1}{\sqrt{2}}\left[\int dE \cdot f_{E_0-\Delta E/2}(E)|E,H\rangle + \int dE \cdot f_{E_0+\Delta E/2}(E)|E,V\rangle\right], \quad (1)$$

where $|H\rangle$ and $|V\rangle$ are photon polarization states, and $f_{E_0}(E)$ is the lineshape of the energy level given by the amplitude of a Lorentzian distribution

$$f_{E_0}(E) = \sqrt{\frac{\Gamma}{2\pi}} \frac{1}{(E-E_0)+i\Gamma/2}, \quad (2)$$

It is useful to note that for small splitting compared to linewidth, $\Delta E/\Gamma \ll 1$, one can expand Eq.2 to obtain

$$f_{E_0\pm\Delta E/2}(E) = \sqrt{\frac{\Gamma}{2\pi}} \frac{1}{E-(E_0\pm\Delta E/2)+i\Gamma/2} \approx f_{E_0}(E) \pm \sqrt{\frac{\Gamma}{2\pi}} \frac{\Delta E/2}{(E-E_0+i\Gamma/2)^2} \quad (3)$$

In general, the observed spectrum is a convolution of the spectral response of the grating with the actual photon spectrum, but it is convenient to consider the limit of an ideal spectrometer.

We consider post-selecting the polarization of the photon emitted from the exciton level in the state $|f\rangle = \frac{1}{\sqrt{2}}\left[(1-\delta)|H\rangle - (1+\delta)|V\rangle\right]$, whose overlap with the state $|i\rangle$ has a magnitude $\delta$, which we assume to be real and much smaller than 1. This projects the energy onto the state

$$|\Psi_p\rangle_{\Delta E} = \frac{1}{2\sqrt{P}}\left[(1-\delta)\int dE \cdot f_{E_0-\Delta E/2}(E)|E\rangle - (1+\delta)\int dE \cdot f_{E_0+\Delta E/2}(E)|E\rangle\right], \quad (4)$$

where $P = \delta^2 + \Delta E^2/2\Gamma^2$ is the post-selection success probability in the limit of $\Delta E/\Gamma \ll 1$. The post-selection can succeed because of the finite $\delta$, or because of the phase shift picked up by the spin state as a result of energy difference, proportional to $\Delta E/\Gamma$. In order to obtain the average energy shift we restrict ourselves to the former case and will later discuss the contribution of the second term to limitations imposed on the amplification. We can now evaluate the expectation value of the energy, using the state from Eq. 4 and the lineshape from Eq. 3

$$\left\langle \Psi_p \left| \hat{E} \right| \Psi_p \right\rangle_{\Delta E} \approx E_0 + \frac{\Delta E}{2\delta} \equiv E_0 + \Delta E_w, \tag{5}$$

where $\hat{E}$ is an operator measuring the energy. The energy operator in our scheme is an ideal projection operator. In experimental realizations of the proposal, the energy spectrum is measured with finite resolution, however for WVA larger than spectral resolution it should not affect the amplification. The average value of the energy shift, $\Delta E_w$, in this post-selected distribution is the weak value in our scheme and it can be much larger than the actual splitting in the initial distribution (Fig. 1 c).

The formalism described above neglects the finite lifetime of the exciton state, assuming that as $\delta$ vanishes, the post-selection never succeeds. However, the post-selection can still succeed due to the accumulated-phase uncertainty as a result of the uncertain energy difference of the two spin states and since WVA relies on the interference in the probe state, this effective dephasing makes the WVA less significant (Fig. 2). Therefore, for any given $\Delta E$ there is an optimum $\delta$ for achieving the largest WVA. To first order, this occurs when the post-selection has equal chances of succeeding due to overlap or due to accumulated phase, which can be shown to occur at $\delta_{opt} \approx \Delta E / (\sqrt{2}\, \Gamma)$. The maximum value of the amplification is approximately $\Gamma / (2\sqrt{2} \Delta E)$, which means that the shift may be enhanced up to a value on the order of the linewidth $\Gamma$. The post-selection parameter $\delta$ is crucial in the WVA [9, 15,16]. In our scheme, $\delta$ is the cosine of the angle between the polarization of the first photon on the $XX^0$ line and polarization of the second photon on the $X^0$ line. Both detected polarizations can be determined experimentally using polarizers (Fig. 1 a, b). The smallest values of $\delta$ discussed in our paper correspond to polarizer rotation on the order of 0.1 degrees, which is feasible with practical experimental components.

Despite the significant amplification in WVA, it is accompanied by a reduction in the sample size due to post-selection. In the case of a fermionic system such as a QD, the lifetime of the levels sets a fundamental limit on the highest rate of the QD-based emitters due to the PEP allowing only one fermion per state. Therefore, slow noise in fermionic systems makes a significant contribution. If a QD is being pumped at a rate $R_p$, the maximum possible SNR in this system is determined by how fast the quantum dot can be reloaded – limited by the lifetime of the excited state, $R_{p,\max} = 1/T_1$ (Fig. 3, green dashed line). However, the presence of any noise with correlation time slower than $T_1$ can lower the highest possible SNR even more (Fig. 3, blue solid line). In Fig. 3, the noise is modeled by an exponential correlation function with a time constant $\tau_c$; there is therefore a knee in the SNR when the measurement rate equals $1/\tau_c$. The post-selection makes the relevant measurement events less frequent and therefore the noise corresponding to those events is less correlated. Therefore, it is beneficial to increase the SNR by WVA beyond the limit set by the slow noise (Fig. 3, red dashed-dotted line). In general, the SNR enhancement due to WVA depends on various parameters of the system such as noise characteristics, post-selection parameter, and lifetime. In the specific case presented on Fig. 3, for equal noise characteristics and photon emission rates, the conventional measurement SNR=15, whereas with WVA the enhanced SNR=50. Thus the enhancement of SNR due to WVA in this specific case is more than threefold, and can be larger for different system parameters. Similar to the pointer shift (Fig. 1c), the optimal SNR (Fig 3 inset) is obtained when $\delta_{opt} \approx \Delta E / (\sqrt{2}\,\Gamma)$.

Another important factor to be taken into account for WVA, and for solid-state systems in particular, is the dephasing rate of the probe. Different dephasing processes can result in loss of purity in the probe state and since the WVA occurs due to interference in the probe, dephasing results in reduced amplification. However, since the amplification relies on interference between energy components of the electron wavefunction with different spin states, only noise processes which are spin-dependent and affect the relative phase of the two spin states can result in suppressed amplification. In the presence of such processes the state of the H-polarized photons can be written as

$$\rho = \int dE_{noise} P_{noise}(E_{noise}) |\Psi_p\rangle_{\Delta E + E_{noise}} \langle\Psi_p|_{\Delta E + E_{noise}}, \tag{6}$$

where the dephasing is modeled by random energy shifts with a probability distribution $P_{noise}(E_{noise})$. As an example we consider a Lorentzian distribution with the width $\gamma$ centered at $E_{noise} = 0$. For a given overall emission width, as the ratio $\gamma/\Gamma$ increases, the purity of the probe decreases, and one expects smaller amplification. Fig. 4 shows that as the contribution of the dephasing, $\gamma/\Gamma$, increases, the WVA decreases. Increasing $\gamma$ pushes the optimum $\delta$ to larger values, limiting the achievable amplification. It is also important to note that the lifetime of the exciton state determines which portion of the noise spectrum can degrade the WVA: if the timescale of the noise is longer than the exciton lifetime, the spectral components of the emitted photon from the two spin states remain in phase long enough to interfere. The optimal amplification is reduced significantly when dephasing becomes comparable to $\Delta E$ (Fig.4 inset). This occurs because of the interferometric nature of WVA, which is strongly degraded by dephasing larger than $\Delta E$.

The proposed WVA should significantly increase the resolution of different kinds of spectroscopy in principle. The main goal of spectroscopy is to resolve various interaction energies. Our approach allows taking the WVA idea to practical applications by significantly increasing the SNR, and thus the resolving capability, of any technique aimed at studying interaction energies limited by slow noise. Moreover, our proposal is applied here for the first time to the study of interactions between fermions. Fermions are inherently prone to slow noise because of the PEP, and their interaction has to be probed one particle at a time – unlike bosons which can be prepared in a state with a large number of bosons populating a single mode. Various fermion-fermion interactions with the energy splitting smaller than the natural linewidth can be probed by the proposed WVA method. In addition to electron-hole interactions in semiconductor structures, WVA can be employed in spectroscopy of various systems such as dipole-dipole spin-dependent interaction in fermionic cold atoms [17,18] and in bosonic systems composed of fermions including cold-atom Bose Einstein condensates [19] as well as in solid-state systems such as exciton-polariton condensates in semiconductor strongly coupled microcavities [20]. The interaction studied in our paper is between two fermions, as a specific example: anisotropic electron-hole exchange interaction in QDs. The proposed scheme employs photons to carry information; however, the use of photons here is only a convenient tool to prepare the initial state and to read out the final state. Optical techniques are in general very useful for access to information in the studies of both bosons and fermions. Almost all fermion studying techniques employ photons as a tool including solid-state pump-probe experiments [21], cold atom interaction experiments such as strongly interacting Rydberg atoms [22], and the study of the interaction between an electron and a nucleus in an atom resulting in the atomic levels including the fine and the hyperfine structure [23] – all use photons to carry information. Nevertheless, the physical effect at the core of this study, and the potential future studies is an interaction between fermions, which can be resolved only with WVA.

In conclusion, we have proposed a scheme for enhanced probing of a few-fermion interaction and analyzed it for a specific case of electron-hole exchange interaction by mapping the QD state, including energy and spin, into a photon. We study the effect of noise and dephasing on the WVA. Such post-selection-enhanced probing of the interactions can significantly improve the SNR in studies of few-fermion systems in the presence of slow noise.

A.H. and A.F. contributed equally to this work. We would like to acknowledge financial support from the Natural Sciences and Engineering Research Council of Canada and the Canadian Institute for Advanced Research.

**References:**

[1] L. A. Rozema, A. Darabi, D. H. Mahler, A. Hayat, Y. Soudagar, A. M. Steinberg, Phys. Rev. Lett. **109**, 100404 (2012).

[2] S. Kocsis, B. Braverman, S. Ravets, M. J. Stevens, R. P. Mirin, L. K. Shalm, and A. M. Steinberg, Science **332**, 1170 (2011).

[3] K.J. Resch, J.S. Lundeen, A.M. Steinberg, Phys. Lett. A **324**, 125 (2004).

[4] J. S. Lundeen and A. M. Steinberg, Phys. Rev. Lett. **102**, 020404 (2009).

[5] K. Yokota, T. Yamamoto, M. Koashi and N. Imoto, New J. Phys. **11**, 033011 (2009).

[6] O. Hosten and P. Kwiat, Science **319**, 787 (2008).

[7] P. B. Dixon, D. J. Starling, A. N. Jordan, and J. C. Howell, Phys. Rev. Lett. **102**, 173601 (2009).

[8] Y. Aharonov, D. Z. Albert, and L. Vaidman, Phys. Rev. Lett**. 60**, 1351 (1988).

[9] A. Feizpour, X. Xing, and A. M. Steinberg, Phys. Rev. Lett. **107**, 133603 (2011).

[10] H. Kosaka, T. Inagaki, Y. Rikitake, H. Imamura, Y. Mitsumori and K. Edamatsu, Nature **457**, 702 (2009).

[11] C. Santori, D. Fattal, M. Pelton, G. S. Solomon, and Y. Yamamoto., Phys. Rev. B **66**, 045308 (2002).

[12] N. Akopian, N. H. Lindner, E. Poem, Y. Berlatzky, J. Avron, D. Gershoni, B. D. Gerardot, P. M. Petroff, Phys. Rev. Lett. **96**, 103501 (2006); R. M. Stevenson, R. J. Young, P. Atkinson, K. Cooper, D. A. Ritchie, A. J. Shields, Nature **439**, 179 (2006);

[13] J. Berezovsky, M. H. Mikkelsen, N. G. Stoltz, L. A. Coldren, D. D. Awschalom, Science **320**, 349 (2008).

[14] T. Warming, E. Siebert, A. Schliwa, E. Stock, R. Zimmermann, and D. Bimberg, Phys. Rev. B **79**, 125316 (2009); Y. Benny, Y. Kodriano, E. Poem, S. Khatsevitch, D. Gershoni and P. M. Petroff, Phys. Rev. B **84**, 075473 (2011).

[15] Y. Ota, S. Ashhab and F. Nori, Phys. Rev. A **85**, 043808 (2012).

[16] M. Iinuma, Y. Suzuki, G. Taguchi, Y. Kadoya and H. F Hofmann, New J. Phys. **13**, 033041 (2011).

[17] P. Wang, Z.-Q. Yu, Z. Fu, J. Miao, L. Huang, S. Chai, H. Zhai, and J. Zhang, Phys. Rev. Lett. **109**, 095301 (2012).

[18] L. W. Cheuk, A. T. Sommer, Z. Hadzibabic, T. Yefsah, W. S. Bakr, and M. W. Zwierlein, Phys. Rev. Lett. **109**, 095302 (2012).

[19] A. Widera, F. Gerbier, S. Fölling, T. Gericke, O. Mandel and I. Bloch, New J. Phys. **8**, 152 (2006).

[20] C. Schneider, A. Rahimi-Iman, N. Y. Kim, J.Fischer, I. G. Savenko, M. Amthor, M. Lermer, A. Wolf, L. Worschech, V. D. Kulakovskii, I. A. Shelykh, M. Kamp, S. Reitzenstein, A. Forchel, Y. Yamamoto and S. Höfling, Nature **497**, 348 (2013).

[21] J. A. Kash, Phys. Rev. B **40**, 3455–3458 (1989).

[22] K. Afrousheh, P. Bohlouli-Zanjani, D. Vagale, A. Mugford, M. Fedorov, and J. D. D. Martin, Phys. Rev. Lett**. 93,** 233001 (2004).

[23] A. Cingoz, A. Lapierre, A.-T. Nguyen, N. Leefer, D. Budker, S. K. Lamoreaux, and J. R. Torgerson, Phys. Rev. Lett**. 98,** 040801 (2007).

**Figure captions:**

**Figure 1.** (color online). (a). The initial spin state in the QD level is prepared by selecting the polarization of the photon emitted in bi-exciton to exciton decay. (b) After polarization post-selection, the spectrum of the photon from exciton decay is observed. (c) calculated post-selected probe spectra vs. post-selection parameter $\delta$. The white dashed line is the average probe shift for each $\delta$. The vertical dash-dotted lines correspond to the average energy of the two exciton states. The dotted line shows the idealized weak value, i.e. the first-order approximation shift value calculated in Eq. 5.

**Figure 2.** (color online) Calculated weak-value enhanced probe shift versus post-selection parameter $\delta$ and energy splitting, $\Delta E$.

**Figure 3.** (color online) Calculated SNR versus QD pumping rate in the presence of slow noise with time constant $\tau_c$. The dashed green line shows the maximum possible SNR determined by the maximum reloading time of the QD, $\Gamma = 1/T_1$. The solid blue line shows the degraded SNR in the presence of noise with timescale $\tau_c$. The dotted-dashed red line is the enhanced SNR due to post-selection. The inset is the SNR dependence on post-selection parameter, $\delta$, for high pump rates and for $\Delta E=0.1\Gamma$.

**Figure 4.** (color online) Calculated probe shift versus dephasing rate $\gamma$ and post-selection parameter, $\delta$. The dotted line shows the optimum probe shift for different values of the dephasing rate. The inset is the calculated optimal amplification vs. $\gamma/\Delta E$ for different values of $\Delta E$.

**Figure 1**

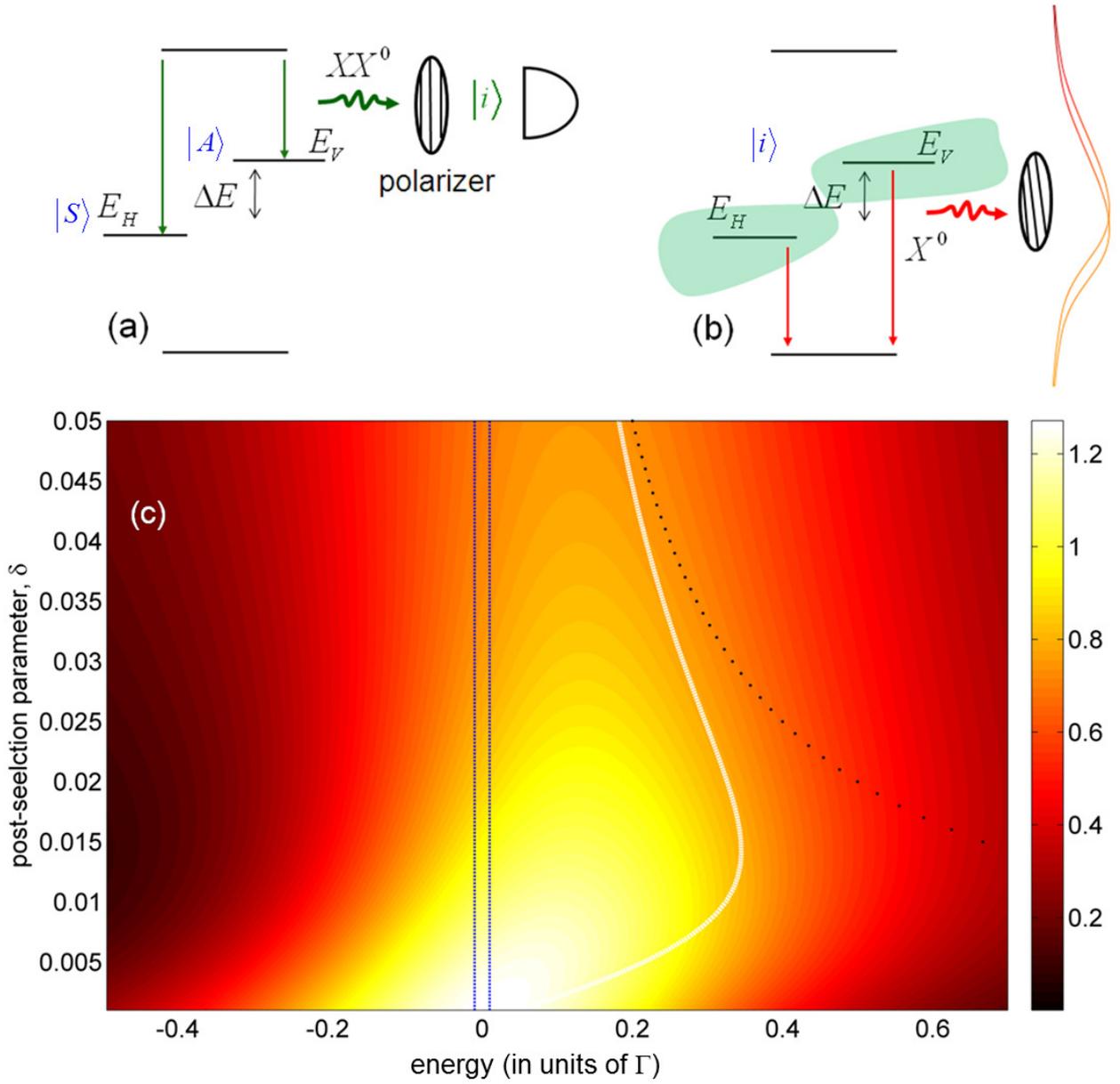

**Figure 2**

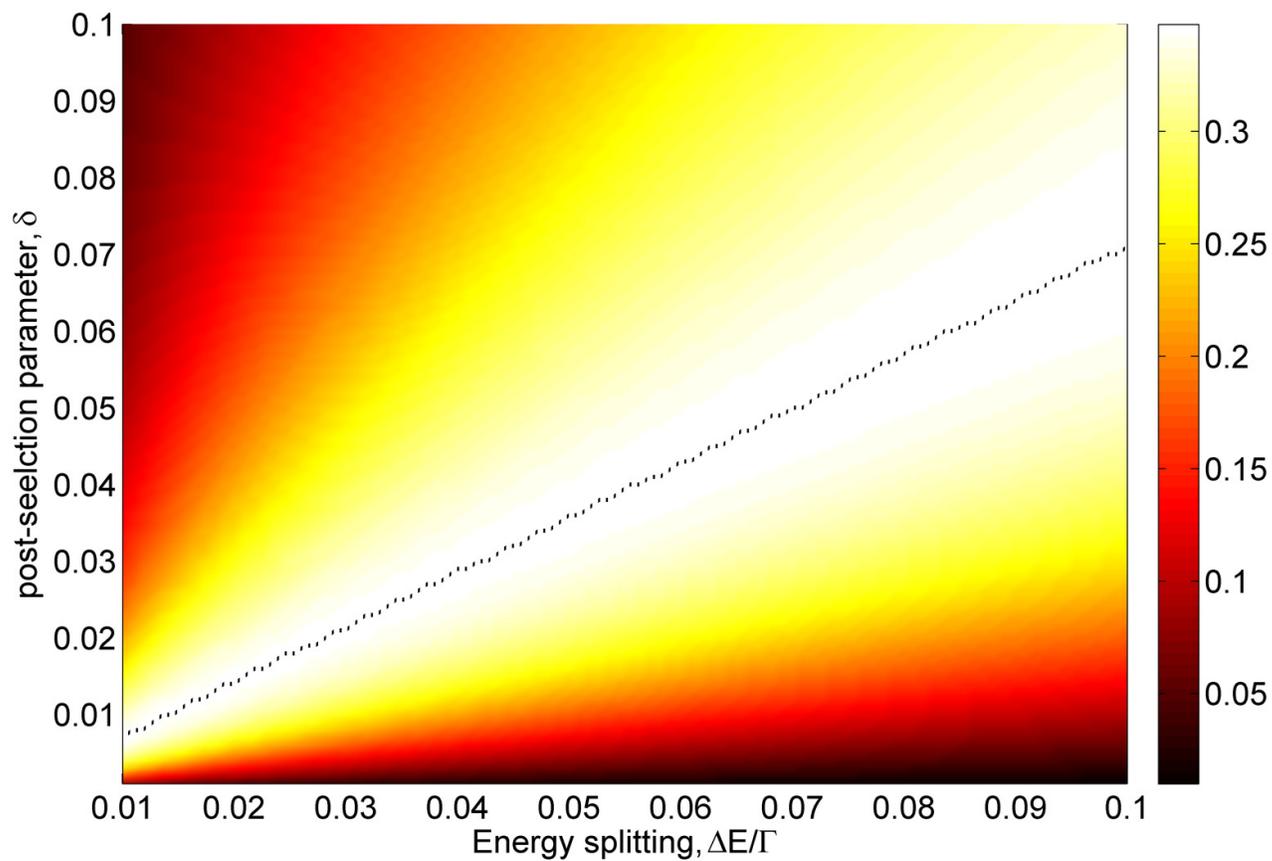

**Figure 3**

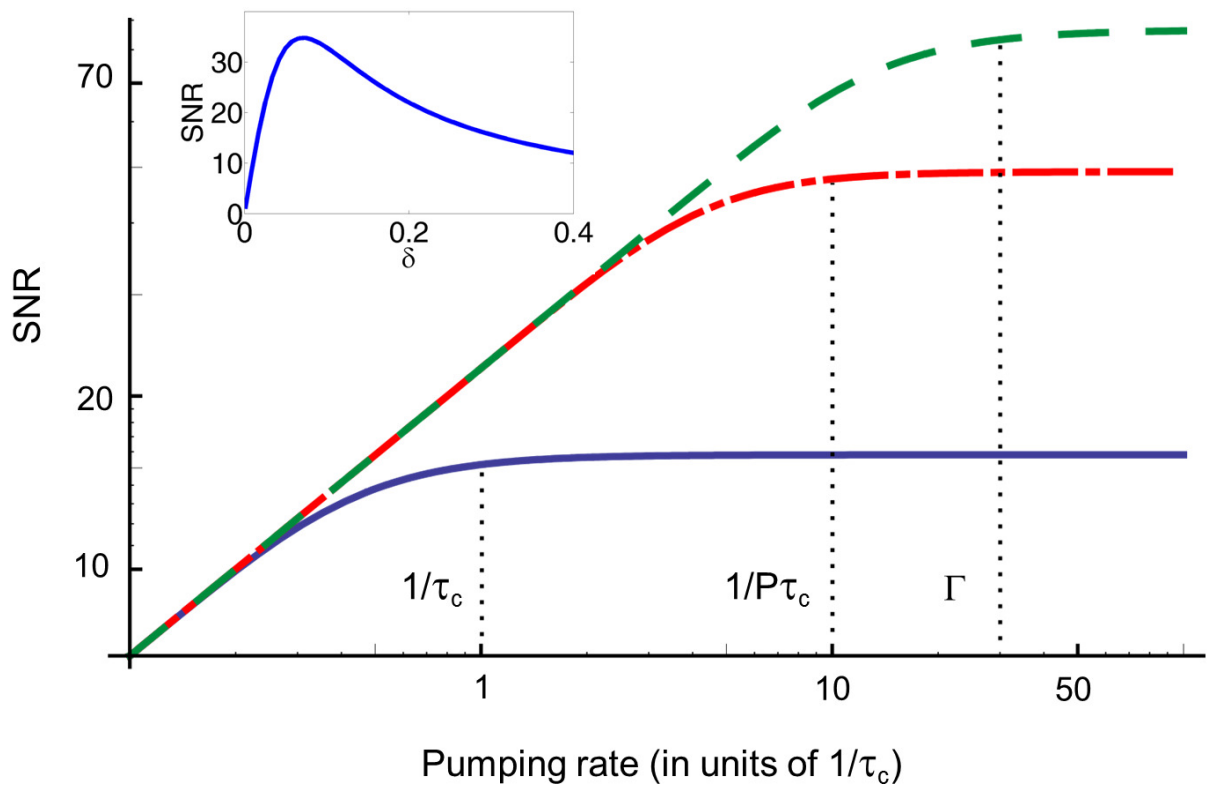

**Figure 4**

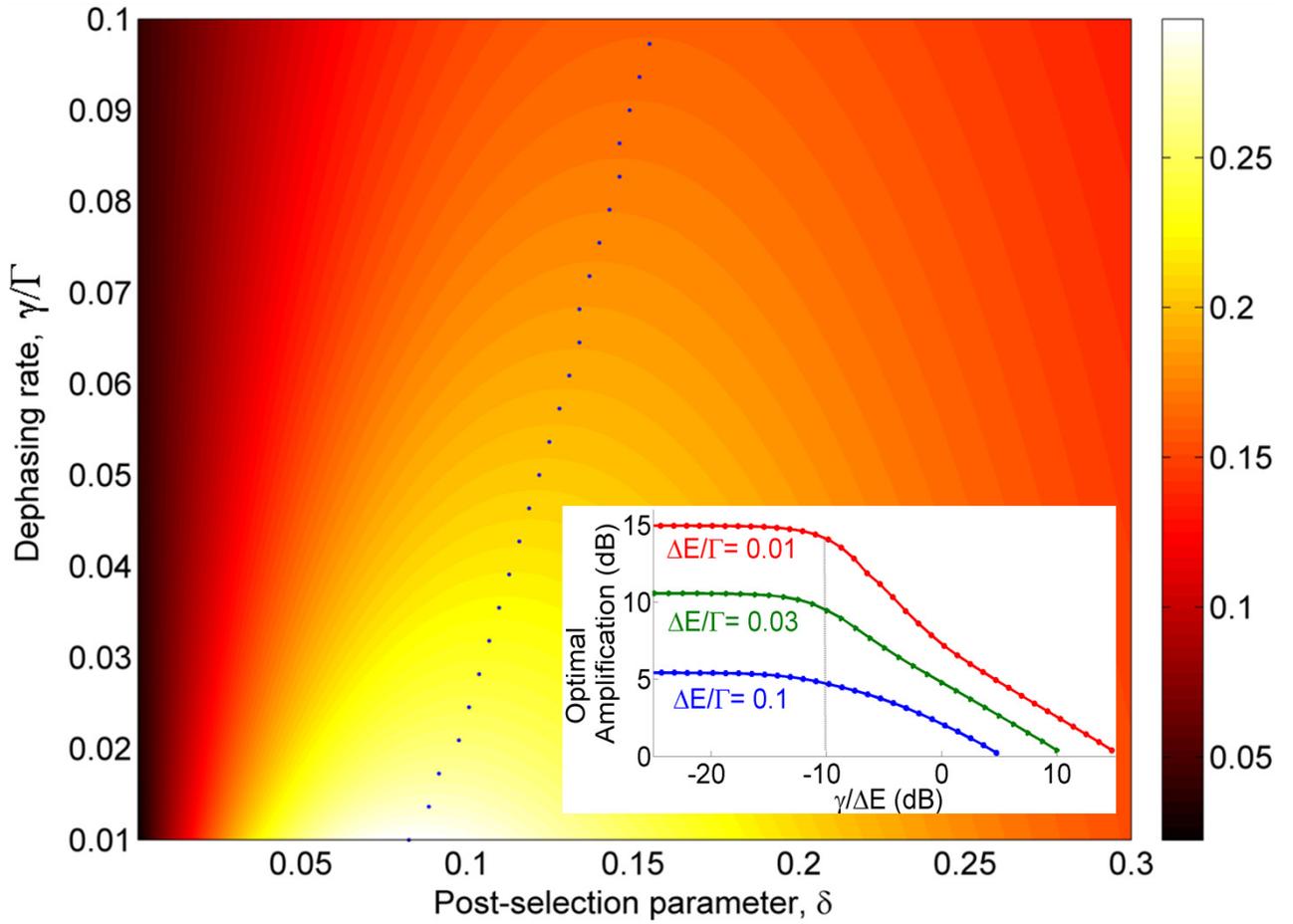